# A Low-Cost Robot Science Kit for Education with Symbolic Regression for Hypothesis Discovery and Validation

**Authors:** Logan Saar[1], Haotong Liang[1], Alex Wang[1], Austin McDannald[2], Efrain Rodriguez[3], Ichiro Takeuchi[2], A. Gilad Kusne[1,2,*]

**Affiliations:**
[1] Materials Science and Engineering Department, University of Maryland, College Park, MD 20742, US
[2] Materials Measurement Science Division, National Institute of Standards and Technology, Gaithersburg, MD 20899, US
3 Chemistry and Biochemistry Department, University of Maryland, College Park, MD 20742, US
* Correspondence: aaron.kusne@nist.gov

## Abstract
The next generation of physical science involves robot scientists – autonomous physical science systems capable of experimental design, execution, and analysis in a closed loop. Such systems have shown real-world success for scientific exploration and discovery, including the first discovery of a best-in-class material. To build and use these systems, the next generation workforce requires expertise in diverse areas including ML, control systems, measurement science, materials synthesis, decision theory, among others. However, education is lagging. Educators need a low-cost, easy-to-use platform to teach the required skills. Industry can also use such a platform for developing and evaluating autonomous physical science methodologies. We present the next generation in science education, a Low-cost Autonomous Scientist kit. The kit was used during two courses at the University of Maryland to teach undergraduate and graduate students autonomous physical science. We discuss its use in the course and its greater capability to teach the dual tasks of autonomous hypothesis (i.e., model) exploration, optimization, and determination, with an example of autonomous experimental "discovery" of the Henderson–Hasselbalch equation.

## Introduction

The need for robotic science is growing rapidly, as exemplified by a central challenge of materials discovery. Advances in technology often require better materials. However, scientists are quickly exhausting the materials that are simpler to make, e.g., materials of simple stoichiometry (few elements) and few processing steps. As a result, scientists are driven to explore materials of greater complexity. With each new synthesis or processing parameter, the number of possible materials grows exponentially. This growing number of materials holds great promise but comes with a significant challenge – a rapidly growing number of materials to explore. The traditional Edisonian search for better materials consists of one-by-one materials synthesis, characterization, and data analysis. This approach to materials discovery becomes infeasible as the search space grows. High-throughput methods[1] and machine learning (ML) make it possible to synthesize, characterize, and analyze hundreds of materials in days, but this geometric speedup can't keep up with the exponential challenge.

The most recent innovation in the search for advanced materials is the use of active learning – the optimal experiment design field of ML. Active learning-based recommendation engines guide experiments both in the lab and in silico, accelerating the discovery of novel materials. By integrating active learning with automated tools, closed loop autonomous physical science (APS) systems - i.e., robot scientists[2] - become possible. For these systems, each subsequent materials experiment is selected and executed to maximize knowledge for the user. For example, APS systems have been used in biology to optimize synthesis routes for yeast enzymes[3] and in chemistry to identify thin film molecular mixtures with improved photoactivity[4]. Additionally, for the first time an APS system has discovered a best-in-class solid state material – the new best-in-class phase change memory material[5]. Autonomous techniques have also advanced measurement science, accelerating X-ray[5,6] and neutron diffraction[7] at multiple user facilities. Diverse companies now seek to use APS in their research and development pipelines.

A next-generation-workforce is needed to fuel the growth of APS. However, with the rapid development of APS, educators have been left behind. Universities are scrambling to integrate next-generation-workforce data science skills into their courses[8] in traditional non-data-centric disciplines. Scientific societies have begun designing dedicated machine learning tutorials at workshops and conferences. Some stand-alone resources exist. The authors of this article run a bootcamp –the Machine Learning for Materials Research[9]. The five-day, hands-on bootcamp is now on its seventh year, teaching programming and ML fundamentals through APS skills to students, academic, national lab, and industry scientists. The authors also host the REsource for Materials Informatics[10] (REMI) website which curates online materials informatics and APS tools including data science Jupyter notebooks.

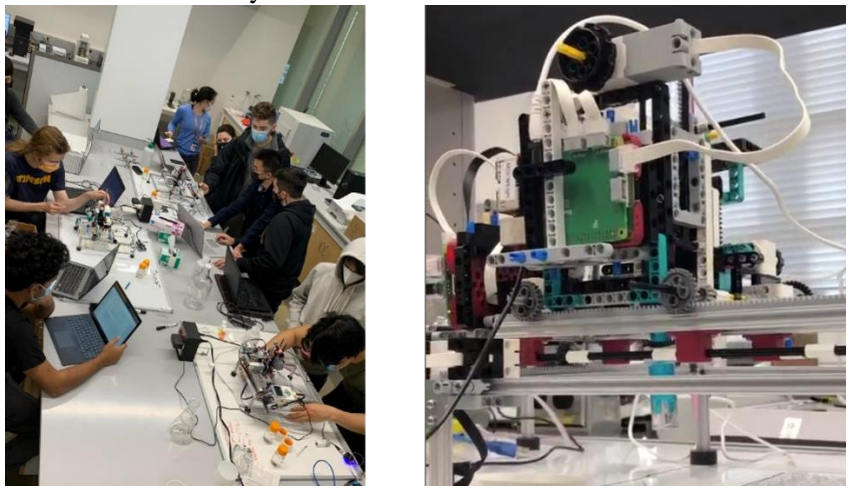

Figure 1. Left) Image of student projects from University of Maryland course in machine learning for materials science. Students were split into groups and each group worked with a robot scientist. Right) Image of the robot scientist with a pH sensor.

While the list of education resources is growing, there is something lacking – a low-cost, easy-to-use APS platform on which APS can be learned, demonstrated, and explored. There is a growing number of commercial APS systems, but they cost hundreds of thousands of dollars. Systems demonstrated in the literature are costly, bespoke, and involve highly complex equipment, making them inappropriate for the classroom. R Deneault, et al. demonstrated a potential educational tool – a closed-loop autonomous 3D printer, capable of autonomously optimizing the printed line profile, with a preliminary cost of over $2,500[11]. The authors also discuss the future aim to build a lower cost version for demonstrating physical science experiments. An education platform must be lower-cost, robust, and easy-to-use for teaching the many needed APS skills. Such a system can also serve industry for APS methods development.

The list of APS skills required of the next generation workforce is long. Physical science skills include experimental design, materials synthesis, characterization, and analysis. Data science skills are needed to ensure that collected data is findable, accessible, interoperable, and reuseable[12]. A next-generation scientist will need to know the basics of software and hardware design, such as designing for safety when dealing with toxic or delicate materials. Control systems knowledge is required to manage APS motion and environmental controls. ML knowledge is required for closed-loop data analysis, prediction, and experiment design. Knowledge of uncertainty quantification and propagation is also key, as proper uncertainty handling improves ML performance including active learning-based experiment design.

Scientific ML[13] – also known as inductive bias ML, is another key knowledge domain for the next-generation workforce. Many physical science challenges suffer from a sparsity of data despite having an abundance of prior knowledge encoded in physical laws and heuristics. Scientific ML incorporates this prior knowledge into the ML pipeline to achieve physically meaningful analysis, prediction, and experiment design with greater performance. As a result, scientific ML can greatly reduce the number of APS experiments needed to achieve a user-defined goal. An optimal APS education platform should allow students and professionals to learn and develop novel scientific machine learning algorithms.

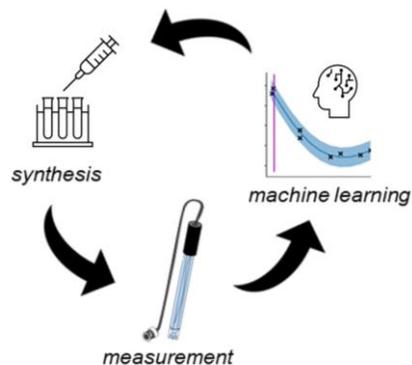

Figure 2. Image of the applied physical science cycle used in the experiment. Samples are synthesized by mixing acid and base in a well, a sensor measures pH, the data is analyzed, and the next experiment is chosen.

We present the next generation in science education and APS methodology development, a Low-cost Autonomous Science platform. The robot scientist kit is an easy-to-use, modular system based on modular toy parts, 3D printed parts, Raspberry pi components and aluminum extrusions with a total cost of less than $300. The system was inspired by the work of Gerber et al.[14] to build an education set for teaching chemistry. The kit was used to teach undergraduates and graduate students during two machine learning courses at the University of Maryland. Students first learned the fundamentals of APS through lectures and hand-on exercise with Jupyter notebooks. They were then split into groups to work hands-on with the kits to solve challenges of APS exploration and discovery. The robot scientist was adapted for liquid handling and pH testing, allowing the students to apply their APS knowledge to investigate the relationship between acid-base ratio and the resulting pH, a relationship described (within a certain mixture range) by the Henderson–Hasselbalch (HH) equation. The acid and base used are on the level of vinegar and milk of magnesia, respectively, so the following experiments can be performed with food safe, household liquids. The students were also provided template APS code, which they were asked to modify to solve the challenges.

Figure 1 shows a photo from the course and a photo of the robot scientist with a pH sensor. Figure 2 presents the system workflow used by the students. ML is used to analyze previous data and active learning is used to determine the next acid-base ratio to investigate. Automated liquid handling creates the desired sample, and automated sensing measures the sample pH. The new data is then used to update the ML model and guide subsequent experiment design, i.e., mixing ratio of acid to base. In this work we discuss the course-based projects provided to the students for synthesis-property relationship exploration and optimization. We also present more advanced uses for the kit including on-the-fly hypothesis design and validation using probabilistic physical models, i.e., identifying the mechanistic rule underlying the studied relationship. Alternatively, for a simpler challenge for younger students, the system has also been used to autonomously match a given color through a mounted camera and access to water with red, green, and blue food coloring. We aim to make the kit available to the community, along with teaching material and a vibrant software ecosystem. A video of the system performing the closed-loop pH experiments can be seen at: https://youtu.be/TtPM7zXI5kQ

## Discussion

The students were asked to solve two challenges: 1) To write a code and operate the system, so that it can autonomously identify the relationship between acid to base ratio and the resulting pH using the minimum number of experiments. 2) To modify the code so that the system identifies the acid to base ratio with a pH of 4.5 using the minimum number of experiments. They were also introduced to scientific challenges, such as noisy pH sensor measurements. Each student group was asked to write a Jupyter notebook to address the two challenges. Figure 3 shows a solution to the challenges, where off-the-shelf Gaussian process regression (using a Matern 5/2 basis function kernel for its flexibility[15]) was fit to the data and a) paired with an exploratory active learning acquisition function for Challenge 1 and b) paired with an acquisition function that balances exploration and exploitation for Challenge 2. The two acquisition functions recommend different subsequent experiments.

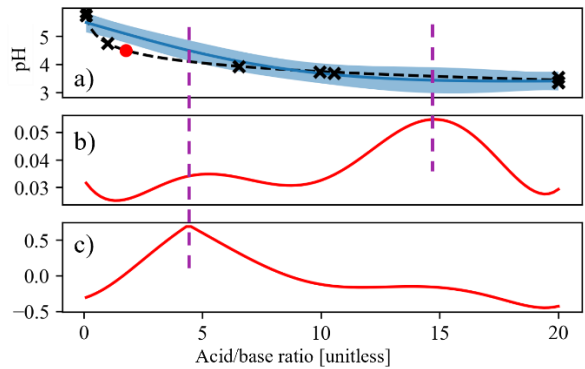

Figure 3. a) Gaussian process fit to experiment data. The hidden true function (black dashed), GP mean (blue line), and GP 95 % confidence interval (blue region) are indicated. b) Exploratory acquisition function for Challenge 1, c) Acquisition function that balances exploration with optimization, a solution for Challenge 2. The next sample to select is indicated by the dashed purple line, with an acid/base ratio of approximately 15 selected for Challenge 1 and approximately 3 for Challenge 2.

At the end of the 3-week project, students presented their results and their APS strategies and discussed opportunities for future improvements. The students found that with APS, they were able to solve the challenges with less than ten experiments, compared to an exhaustive study which takes dozens of experiments. Similarly, total experiment time fell from hours to minutes.

Three advanced challenges were provided to an undergraduate student to solve using the robot scientist. In the first challenge, the student was provided the HH equation (Eqn 1) and asked to use the robot scientist to find the parameter values in the minimum number of experiments. The student used Bayesian inference and active learning to guide subsequent experiments and focus in on the correct parameter values. Bayesian inference uses Bayes rule (Eqn 2) and probabilistic sampling to estimate unknown distributions. Here $D$ is the past data and $M(\theta)$ is the model $M$ with parameters $\theta$. Figure 4a and 4b show an example of the learned distributions over the parameters $\alpha$ and $\beta$, and Figure 4c shows the resulting distribution over the HH model. We have implemented a more advanced version of this was recently implemented for APS neutron scattering to identify magnetic dynamic parameters[7].

$$pH = \alpha + \beta \log x, \quad x = [Acid]/[Base] \qquad Eqn\ 1$$

$$p(M(\theta)|D) = p(D|M(\theta))p(M(\theta))/p(D) \qquad Eqn\ 2$$

The second challenge provided to the student was one of hypothesis or model selection. The student was asked to use the robot scientist to figure out the underlying model and its parameters if a set of possible models was provided. Here again Bayesian inference and active learning were combined to guide subsequent experiments. Each model in the set was fit to the data and the distribution over parameter values was determined. By combining the distribution over the set of possible models, the student was able to identify the sample ratio at which the models most disagree – the ratio at which entropy is maximum. This is the ratio which is most informative in differentiating between the possible models.

Figure 4d shows an example of this model comparison, Figure 4e shows the resulting entropy as a function of ratio, and Figure 4f shows the use of Bayesian information Criteria to select the best candidate given model fit and complexity.

The final challenge given to the undergraduate was to use the robot scientist to determine the mechanistic function if no potential models are given, i.e., search over the space of all possible models (i.e., hypotheses) to determine the best model. Here the student combined symbolic regression with active learning in a closed loop. At each iteration, past data is fit using symbolic regression to identify a set of potential models. These models are quantified by fit quality using mean square error (MSE) and complexity (See Figure 5). From these values a score is computed to rank the models. For each model, the MSE provides an estimate of model uncertainty, based on the validated assumption that the measurement noise is normally distributed and heteroskedastic. The next experiment is then selected using the same entropy selection criteria from the previous exercise. The next experiment is performed, data collected, and the cycle is repeated. An example output is given in Figure 5, where 5 models provide adequate fits to the data. The model with the best score is an extremely close fit to the HH equation.

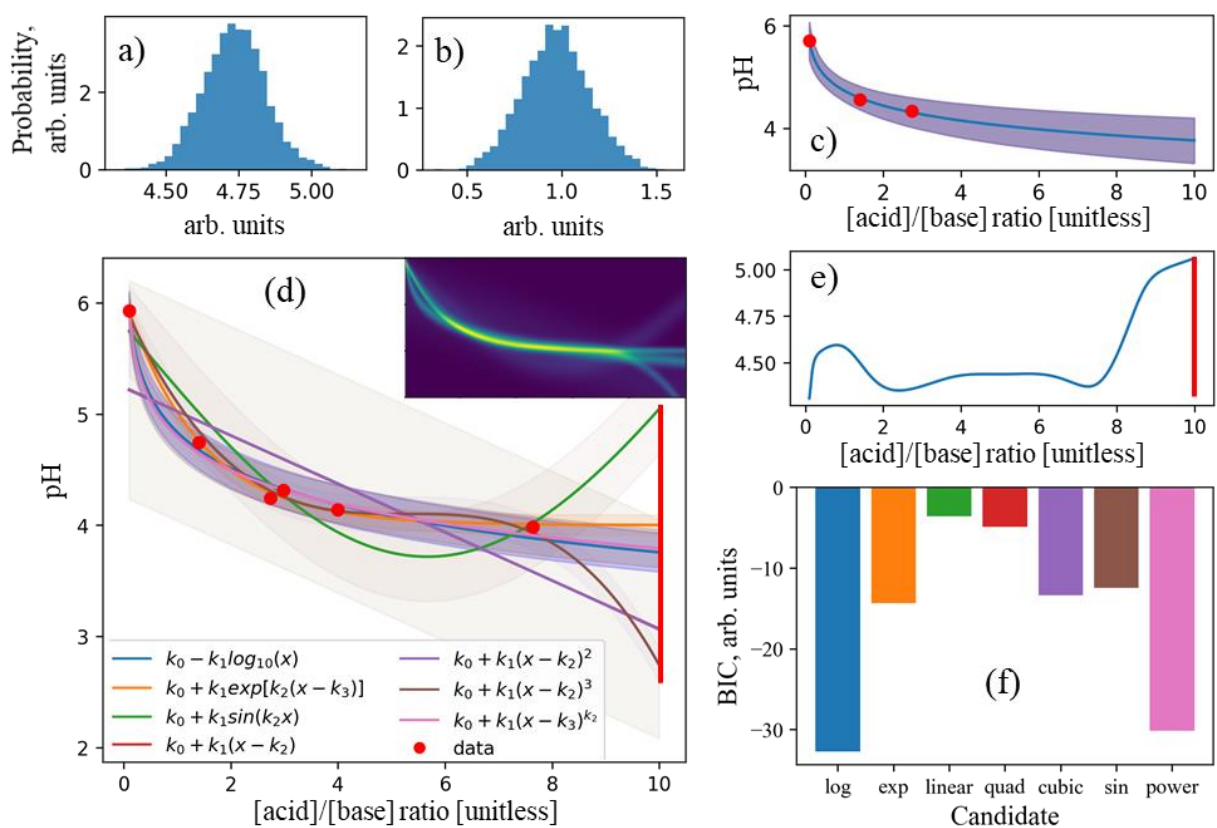

Figure 4. a-c) Bayesian inference and active learning are combined for parameter determination with (a,b) distributions over the parameters and c) the function. (d-f) Bayesian inference and active learning are combined for model and parameter determination with d) distributions over possible models, inset: sum over distributions, e) entropy as a function of ratio, f) Bayesian information criteria for model determination.

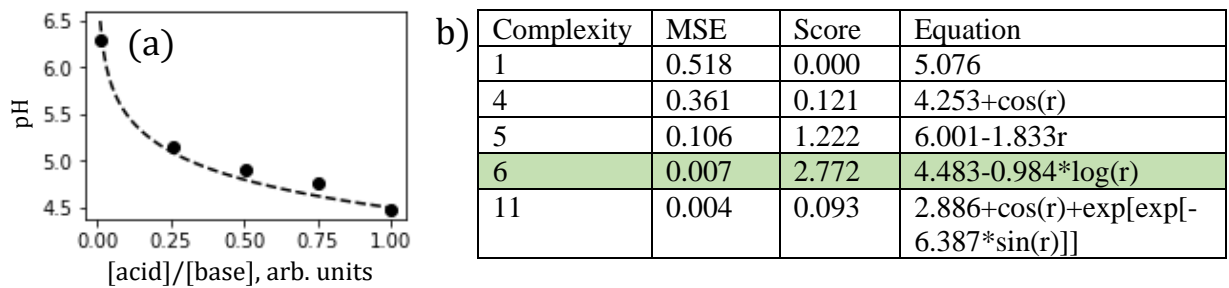

Figure 5. Symbolic regression combined with active learning for probabilistic model determination. a) example data, b) output from symbolic regression with 5 models. The model with the highest score matches the HH equation with a slight deviation of parameters.

## Summary

The robot scientist kit was proven to be an excellent low-cost education platform for teaching APS skills in two courses at the University of Maryland. APS skills taught include automation, closed-loop experimentation, systems control, software design among others. Students learned and executed autonomous ML and scientific ML with exercises in closed loop experiments driven by Gaussian processes and active learning, Bayesian inference for model and parameter determination, and symbolic regression for discovering an unknown mechanistic model and its parameter values. In the process, the students used these skills to learn about color theory (when mixing color water) and chemistry.

During the courses, we identified challenges students face in learning and practicing APS skills, using these lessons to then tune the course lectures, materials, and the robot scientist kit. As a result, we had an excellent student response. Senior undergrads in the course informed us that they were given job offers specifically because of this hands-on course. Industry has also expressed interest in the robot scientist as an APS development platform, comparing it to systems that cost in the hundreds of thousands of dollars. We aim to continue building associated education tools, greater modular functionality, and an open ecosystem for robot scientist users to share their APS code and discoveries. The robot scientist platform will also be incorporated into the previously mentioned Machine Learning for Materials Research bootcamp, expanding the community beyond students to academic, national lab, and industry scientists.

## Conflict of Interest

Ichiro Takeuchi is a founder of MEST, a company working on commercializing the kit.

**Acknowledgements**

A part of this project was supported by and carried out in collaboration with Maryland Energy & Sensor Technologies, LLC.

* **NIST Disclaimer:** Certain commercial equipment, instruments, or materials are identified in this report in order to specify the experimental procedure adequately. Such identification is not intended to imply recommendation or endorsement by the National Institute of Standards and Technology, nor is it intended to imply that the materials or equipment identified are necessarily the best available for the purpose.

## Data Availability
The datasets generated during and/or analyzed during the current study are available from the authors on reasonable request.